\documentclass{frontiersSCNS} 

\usepackage{url,lineno}

\usepackage{natbib}
\usepackage{color}
\usepackage{amsmath}
\usepackage{mathrsfs}
\usepackage{calrsfs}

	\newcommand\ion[2]{#1$\;${\scshape{#2}}}

	\newcommand{\apj}{ApJ}					
	\newcommand{\apjl}{ApJL}				
	\newcommand{\apjs}{ApJS}				
	\newcommand{\aap}{A\&A}				
	\newcommand{\aapr}{A\&A Rev.}			
	\newcommand{\nat}{Nature}				
	\newcommand{\baas}{BAAS}				
	\newcommand{\aspconf}{ASP Conf. Ser.}			
	\newcommand{\procspie}{Proc. SPIE}			
	\newcommand{\ssr}{Space Sci. Rev.}			
	\newcommand{\ao}{Appl. Opt.}				
	\newcommand{\solphys}{Sol. Phys.}			
	\newcommand{\jgr}{JGR}
	\newcommand{\memsai}{MmSAI} 
	\newcommand{\frass}{FrASS}

\def\keyFont{\fontsize{8}{11}\helveticabold }
\def\firstAuthorLast{Raouafi {et~al.}} 
\def\Authors{N. E. Raouafi\,$^{1,*}$, P. Riley\,$^{2}$, S. Gibson\,$^3$, S. Fineschi$^4$, and S. K. Solanki$^{5,6}$}
\def\Address{$^{1}$The John Hopkins University Applied Physics Laboratory, Laurel, MD 20723, USA \\
$^{2}$Predictive Science Inc., San Diego, CA 92121, USA \\
$^{3}$HAO/NCAR, Boulder, CO 80307, USA     \\
$^{4}$INAF $-$ Astrophysical Observatory of Torino, Italy \\   
$^{5}$Max-Planck-Institut f\"ur Sonnensystemforschung, G\"ottingen, Germany \\   
$^{6}$School of Space Research, Kyung Hee University, Yongin, Gyeonggi-Do 446-701, Republic of Korea} 

\def\corrAuthor{N. E. Raouafi}
\def\corrAddress{The John Hopkins University Applied Physics Laboratory, 11100 Johns Hopkins Road
Laurel, MD 20723-6099, USA}
\def\corrEmail{NourEddine.Raouafi@jhuapl.edu}

\begin{document}
\onecolumn
\firstpage{1}

\title[Hanle Effect Diagnostics of the Coronal Magnetic Field]{Diagnostics of Coronal Magnetic Fields Through the Hanle Effect in UV and IR Lines}
\author[\firstAuthorLast ]{\Authors}
\address{}
\correspondance{}
\extraAuth{}
\topic{Coronal Magnetometry}

\maketitle


\begin{abstract}
The plasma thermodynamics in the solar upper atmosphere, particularly in the corona, are dominated by the magnetic field, which controls the flow and dissipation of energy. The relative lack of knowledge of the coronal vector magnetic field is a major handicap for progress in coronal physics. This makes the development of measurement methods of coronal magnetic fields a high priority in solar physics. The Hanle effect in the UV and IR spectral lines is a largely unexplored diagnostic. We use magnetohydrodynamic (MHD) simulations to study the magnitude of the signal to be expected for typical coronal magnetic fields for selected spectral lines in the UV and IR wavelength ranges, namely the \ion{H}{i} Ly-$\alpha$ and the \ion{He}{i} 10830~{\AA} lines. We show that the selected lines are useful for reliable diagnosis of coronal magnetic fields. The results show that the combination of polarization measurements of spectral lines with different sensitivities to the Hanle effect may be most appropriate for deducing coronal magnetic properties from future observations.

 \keyFont{ \section{Keywords:} Sun: corona -- Sun: magnetic fields -- Sun: UV radiation -- Sun: infrared -- Polarization -- Scattering --  Atomic processes -- Plasmas } 
\end{abstract}

\section{Introduction}

Our understanding of coronal phenomena, such as plasma heating and acceleration, particle energization, and explosive activity, faces major hurdles due to the lack of reliable measurements of key parameters such as densities, temperatures, velocities, and particularly magnetic fields. The knowledge of key plasma parameters, particularly the magnetic field and plasma velocity, in the solar corona is a prerequisite to advance our understanding of coronal manifestations that greatly affect and modulate the interplanetary medium, particularly the Earth's environment.

Spectroscopic diagnostics in the ultraviolet (UV) and extreme ultraviolet (EUV) wavelength regimes provide measurements of plasma densities, temperatures, and partial information on the velocity. But they cannot provide any insight into the coronal magnetic field, which is the key player in the structuring of the solar corona and in dominating most (if not all) physical processes underlying the multi-scaled solar activity. For instance, the abundant mechanical energy that is  available in the convection zone is partially transferred to the corona, where it is stored in complex magnetic field structures and dissipated in the form of heat, acceleration, and energization of the plasma during activity events occurring at different spatial and temporal scales. The magnetic activity manifests itself in different forms such as Coronal Mass Ejections [CMEs], flares, jets, waves and instabilities, magnetic reconnection, and turbulence.

Magnetic fields measurements in the photosphere and, to a lesser degree, in the chromosphere, which are based mainly on the Zeeman effect, have been a routine exercise for decades. The Zeeman effect in the higher layers of the solar atmosphere, particularly the corona, are limited to regions of relatively strong magnetic fields (i.e., above active regions) and to infrared lines due to the wavelength squared scaling of this effect and the relatively large widths of coronal lines \citep{2004ApJ...613L.177L}. Other methods based on radio emissions could provide constraints on the magnetic field in the corona \citep[][]{1999SoPh..190..309W,10.3389/fspas.2016.00008}. Coronal magnetic fields are usually approximated through MHD modeling and extrapolation of photospheric measurements \citep[e.g.,][]{2014A&ARv..22...78W}. These approaches have, however, their limitations. For instance, extrapolation models are based on the assumption that the magnetic field is force-free at the lower boundary of the calculation, which is not the case in the photosphere. Additionally, large-scale MHD models of the solar corona are based on synoptic maps of the photospheric magnetic fields, which are built up from images taken by near-Earth observatories recorded over a whole solar rotation. Finally, the MHD models may underestimate the magnetic field strength in the corona \citep{2012JASTP..83....1R}. Moreover, coronal plasma parameters obtained through the models cannot be constrained without direct measurements. For more details on methods for the measurements of magnetic fields in upper solar atmosphere, see reviews by \citet{2001ASPC..248..597F} and \citet{2005ESASP.596E...3R,2011ASPC..437...99R}.

In this paper, we focus on the diagnostics of coronal magnetic fields through the linear polarization of selected spectral lines (i.e., \ion{H}{i} Ly-$\alpha$ and \ion{He}{i} 10830~{\AA}) that are sensitive to the ``Hanle effect". Other spectral lines are also of interest, but the analysis of their polarization is left for future publications.

\section{The Hanle Effect}

\begin{figure}[h!]
\begin{center}
\includegraphics[width=12cm]{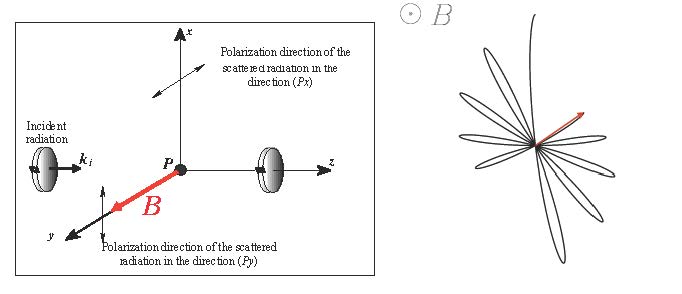}
\end{center}
 \textbf{\refstepcounter{figure}\label{HanleSketch} Figure \arabic{figure}.}{ Illustration of the Hanle effect due to a magnetic field aligned with the line of sight $(Py)$. In the absence of a magnetic field, the direction of the linear polarization of the scattered light is parallel to the $(Px)$ axis (left panel). In the presence of a magnetic field, the combination of the precession around the magnetic field and the damping of the atomic dipole results in a modification of linear polarization that depends on both the strength and direction of the field vector (right panel). }
\end{figure}

The Hanle effect \citep{1924ZPhy...30...93H}, which is the modification of the linear polarization of a spectral line by a local magnetic field, may provide strong diagnostics of regions of weak magnetic fields such as the solar corona, where a number of spectral lines with different but complementary sensitivity ranges are present. Unlike the Zeeman effect, the Hanle effect does not create polarization but requires its presence through other physical processes such as radiation scattering. The Hanle effect is a purely quantum phenomenon and has no classical equivalent. However, for brevity, to provide a simplified illustration of such a complex effect, it can be explained by approximating the excited atom/ion to a damped oscillator with Larmor frequency that is scattering incident non-polarized radiation (see Figure~\ref{HanleSketch}). The Larmor frequency, $\omega_L$, of the precession motion around the magnetic field vector is directly related to the magnetic field strength. The damping is proportional to the finite lifetime, $\tau$, of the upper level of the atomic transition. We note that this classical description could explain only the case of the normal Zeeman triplet (i.e., two-level atom $J_u=1$; $J_l=0$). Significant advances in the theory of radiation scattering in the presence of magnetic fields have been achieved in the last four decades \citep{1977A&A....59..223S,1980A&A....87..109B,1982SoPh...79..291L,1999ApJ...522..524C,2002ApJ...566L..53T,2002ApJ...575..529L,2002A&A...386..721R,2004ASSL..307.....L}.

The sensitivity of a given spectral line to the Hanle effect is a function of the lifetime of the atomic transition and $|{\bf{B}}|$. Ideally, a spectral line is sensitive to magnetic field strengths satisfying the relation 
\begin{equation}
\gamma\ B\ \tau\approx1,
\end{equation}
where $\gamma = g_{J_u}\ \mu_B/\hbar$, $g_{J_u}$ is the Land\'e factor of the upper atomic level, $\mu_B$ is the Bohr magneton, and $\hbar$ is the reduced Planck constant. Practically, the Hanle effect is measurable for $0.2\leq\omega_L\ \tau\leq10$ \citep{1978A&A....69...57B,1982SoPh...78..157B}. Table 1 provides the magnetic field strengths corresponding to the ideal sensitivity of the different spectral lines to the Hanle effect.

\begin{table}[!t]
\textbf{\refstepcounter{table}\label{tbl-1} Table \arabic{table}.}{ Magnetic field strengths corresponding to the ideal Hanle effect for a number of spectral lines (column 4). $A_{ul}$ is the Einstein coefficient of spontaneous emission of the upper level of the corresponding transitions.}
\processtable{}
{\begin{tabular}{lccc}\toprule
Spectral & Wavelength & $A_{ul}$               & $B$     \\
Line       & ({\AA})      & ($10^8$ s$^{-1}$) & (Gauss)  \\
\toprule
\ion{H}{i} Ly-$\alpha$    & \multicolumn{1}{r}{1215.16} & \multicolumn{1}{r}{6.265} & \multicolumn{1}{r}{53.43}  \\
\ion{H}{i} Ly-$\beta$      & \multicolumn{1}{r}{1025.72} & \multicolumn{1}{r}{1.672} & \multicolumn{1}{r}{14.26}  \\
\ion{H}{i} Ly-$\gamma$ &   \multicolumn{1}{r}{972.53} & \multicolumn{1}{r}{0.682} &  \multicolumn{1}{r}{5.81} \\
\ion{H}{i} Ly-$\delta$     &   \multicolumn{1}{r}{949.74} & \multicolumn{1}{r}{0.344} &  \multicolumn{1}{r}{2.93} \\
\ion{O}{vi}                        & \multicolumn{1}{r}{1031.91} & \multicolumn{1}{r}{4.16}   &  \multicolumn{1}{r}{35.48} \\
\ion{He}{i}                      & \multicolumn{1}{r}{10830.0}    & \multicolumn{1}{r}{0.344} &  \multicolumn{1}{r}{0.82}
\\\botrule
\end{tabular}}{}
\end{table}

Theoretically, direct determination of the magnetic field in the solar corona could be achieved through linear polarization of spectral lines with suitable sensitivity to the Hanle effect. The Hanle effect in selected spectral lines yields a powerful diagnostic tool for  magnetic fields typically ranging from a few milli-Gauss to several hundred Gauss (depending strongly on the chosen line and the strength and direction of the magnetic field). Unlike the Zeeman effect, the depolarization of spectral lines by turbulent magnetic fields can be detected in the Hanle regime allowing the determination of the strength of the field in mixed-polarity regions \citep[e.g.,][]{1982SoPh...80..209S,2004Natur.430..326T}.

 In the solar corona, the Hanle effect manifests itself primarily through a depolarization and a rotation of the plane of linear polarization, with respect to the zero-field case where the plane of polarization is parallel to the local solar limb. \citet{1981A&A...100..231B} studied various measurement scenarios allowing for the complete diagnostic of the coronal magnetic field vector. The Hanle effect diagnostic of magnetic fields has been successful in solar prominences \citep{1977A&A....54..811L,1977A&A....59..223S,1980A&A....87..109B,1982SoPh...79..291L,1985SoPh...96..277Q,2002Natur.415..403T,2002ApJ...575..529L}, as well as in arch filament systems \citep{2003Natur.425..692S,2004A&A...414.1109L,2010A&A...520A..77X,2011A&A...532A..63M}.

\subsection{Prominence Magnetic Fields}

\citet{1994SoPh..154..231B} and previous related papers \citep{1977A&A....59..223S,1978A&A....69...57B} have successfully demonstrated the power of the Hanle effect method for measuring the magnetic fields in solar prominences. \citet{1983SoPh...83..135L,1984A&A...131...33L} used observations obtained with the coronagraph polarimeter at the Pic du Midi observatory (France) to study the magnetic field of several hundreds of prominences based on the Hanle effect of spectral lines such as H-$\alpha$, H-$\beta$, and \ion{He}{i} 5876~{\AA}. \citet{1983SoPh...83..135L} found that magnetic field strengths increased with the rise of the solar cycle. They reported an average field strength of $\sim\!6$~Gauss at the beginning of the cycle and about twice this value near solar maximum. Furthermore, \citet{1984A&A...131...33L} found that the magnetic field strength and direction depend also on the prominence height: prominences with heights lower than 30~Mm have $\sim\!20$~Gauss fields with $\alpha\sim20^\circ$ and prominences higher than 30~Mm have $5-10$~Gauss fields with $\alpha\sim25^\circ$ ($\alpha$ is the angle between the magnetic field vector and the prominence spine).

More recently, the \ion{He}{i} 10830 \AA\ triplet has provided additional, very detailed diagnostics of the magnetic field in filaments. Thus, \citet{2009A&A...501.1113K,2012A&A...539A.131K}, Sasso et al. (2011), \citet{2012ApJ...749..138X} found that active region filaments have hectoGauss field strengths, i.e. an order of magnitude larger than the quiet filaments and prominences studied earlier. The spectropolarimetry of this set of lines even revealed the complex multicomponent structure of an activated filament, with the different components displaying magnetic vectors with different field strengths and directions and gas flowing at different speeds and in different directions \citep{2014A&A...561A..98S}.

\subsection{Polarization of Coronal Forbidden Lines}

\citet{1965AnAp...28..877C} has shown that the direction of polarization of some forbidden lines is expected to be either parallel or perpendicular to the local magnetic field projected onto the plane of the sky. This provides a useful approach to study the orientation (direction) of coronal magnetic fields. No information on the field strength can, however, be obtained from such diagnostics.

The polarization of forbidden lines such as \ion{Fe}{xiv}~530.3~nm and \ion{Fe}{xiii}~1074.7~nm have been studied for more than three decades during solar eclipses and using coronagraph observations \citep[][etc.]{1974psns.coll..254Q,1977ROLun..12..109Q,1976BAAS....8..368Q,1987A&A...178..263A}. The relatively low resolution observations show a striking evidence of a predominant radial orientation of the polarization, found everywhere independently of the phase of the solar cycle, which depicts the direction of the coronal field projected on the plane of the sky \citep[see][]{1982A&A...112..350A,1982A&A...116..248A,1987A&A...178..263A}. This may, however, be attributed to the low resolution of the instruments used in the above studies. In addition due to the van Vleck ambiguity, the magnetic field can also be perpendicular to the direction of the linear polarization. This is likely the case at tops of large coronal loops where the magnetic field is nearly horizontal.

\citet{2001ApJ...558..852H} analyzed intensity and polarization maps with better resolution of the \ion{Fe}{xiii}~1074.7~nm line. They found evidence for two magnetic components in the corona: a non-radial field associated with the large-scale structures known as streamers (with loop-like structures at their base) and a more pervasive radial magnetic field, which corresponds to the open coronal magnetic field. More recent observations from the CoMP telescope \citep{2008SoPh..247..411T} with higher resolutions show significant non-radiality of the coronal magnetic field projected on the plane of the sky. For examples of CoMP observations, see \citet{10.3389/fspas.2016.00008}.

\subsection{Polarization of FUV and EUV Coronal Lines}

Several lines in the far UV (FUV) and EUV wavelength ranges have suitable sensitivity to determine the coronal magnetic field via the Hanle effect. The coronal Hanle effect in the FUV and EUV wavelength ranges is largely unexplored despite the high potential of this diagnostic. Li-like ion lines (\ion{O}{vi}, \ion{N}{v}, \ion{C}{iv},...) are presumed to be observed high in the corona due to their broad abundance curves \citep{1986A&A...168..284S}. Li-like ion lines are very intense lines of the chromosphere-corona transition region (the \ion{O}{vi} 103.2 nm line is one of the most intense lines after \ion{H}{i}~Ly-$\alpha$). For instance, the observed emission of the \ion{O}{vi} ion by \citet{1980SoPh...68..187V} at $30^{\prime\prime}$ above the limb \citep[as well as][]{1970ApJS...21....1R} has shown that the \ion{O}{vi} emission extends out into the corona to a few {\it{arcmin}} above the limb. Observations from the Ultraviolet Coronagraph Spectrometer \citep[UVCS;][]{1995SoPh..162..313K} on the The Solar and Heliospheric Observatory \citep[SOHO;][]{1995SoPh..162....1D} show that the \ion{O}{vi} emission extends several solar radii above the limb, with a line ratio and line widths sensitive to Doppler dimming and anisotropic velocity distributions \citep{1998ApJ...501L.127K,1998ApJ...501L.133L}. Such lines have small natural widths and short lifetimes of the upper levels of the corresponding atomic transitions. The magnetic field strength corresponding to their sensitivity to the Hanle effect ranges from a few Gauss to more than 300 Gauss. This interval contains the expected magnitude of the magnetic field strength in the solar corona. 

In a series of papers, \citet{1991OptEn..30.1161F,1993MmSAI..64..441F} and \citet{1995sowi.confR..68F} studied both theoretically and from an instrumental point of view the feasibility of coronal magnetic field diagnostics through the Hanle effect. In particular, they considered the case of the strongest UV coronal line, \ion{H}{i} Ly-$\alpha$. One of the advantages of using \ion{H}{i} Ly-$\alpha$ is the very broad line profile of transition region incident radiation, which makes it insensitive to effects of the solar wind velocity at low coronal heights. Additionally, \ion{H}{i} Ly-$\alpha$ has a negligible collisional component compared to that of the other  \ion{H}{i} Lyman series. This results in a \ion{H}{i} Ly-$\alpha$ zero-field polarization larger than that of the other \ion{H}{i} Lyman lines, increasing the overall line sensitivity to the Hanle effect \citep[see][]{1999SPIE.3764..147F}.

\citet{1999A&A...345..999R} used spectroscopic observations from the SOHO Solar Ultraviolet Measurements of Emitted Radiation spectrometer \citep[SUMER;][]{1995SoPh..162..189W} to measure the linear polarization of the \ion{O}{vi}~103.2~{nm} line. SUMER calibration before launch shows that the instrument is sensitive to the linear polarization of the observed light \citep{1997ApOpt..36..353H}. The observations were made during the roll manoeuver of the SOHO spacecraft on March 19, 1996, in the southern coronal polar hole at $1.3~R_\odot$. For more details on the observations see \citet{1999A&A...345..999R,1999ASSL..243..349R}. The data show in particular that the plane of polarization has an angle of $\sim\!9^\circ$ with respect to the solar limb. In contrast, the polarization direction is expected to be tangent to the local solar limb in the absence of the magnetic field effect. \citet{1999A&A...345..999R,1999ASSL..243..349R,2002A&A...396.1019R} interpreted these measurements in terms of the Hanle effect due to the coronal magnetic field. They developed models to simulate the observational results and inferred a field strength of $\sim3$~Gauss at $1.3~R_\odot$ above the solar pole. The main result from this work is a clear evidence for the Hanle effect in the strong \ion{O}{vi}~103.2~nm coronal lines. This opens a window for direct diagnostic of the coronal magnetic field by using different FUV-EUV lines with complementary sensitivities to the magnetic field.

\citet{2009ASPC..405..423M} discussed the possibility of mapping the on-disk coronal magnetic fields using forward scattering in permitted lines at EUV wavelengths (e.g., the \ion{Fe}{x}~17.4 nm line).


\section{Simulation Data: Magnetic Field Configuration and Plasma Parameters}

\begin{figure*}[!ht]
\begin{center}
\includegraphics[width=0.45\textwidth]{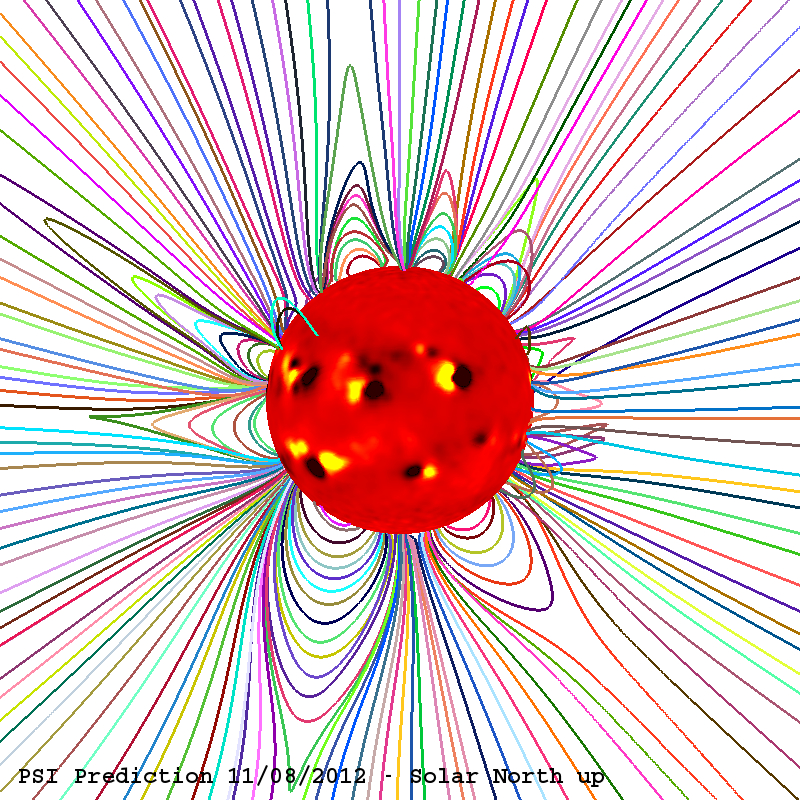}
\includegraphics[width=0.45\textwidth]{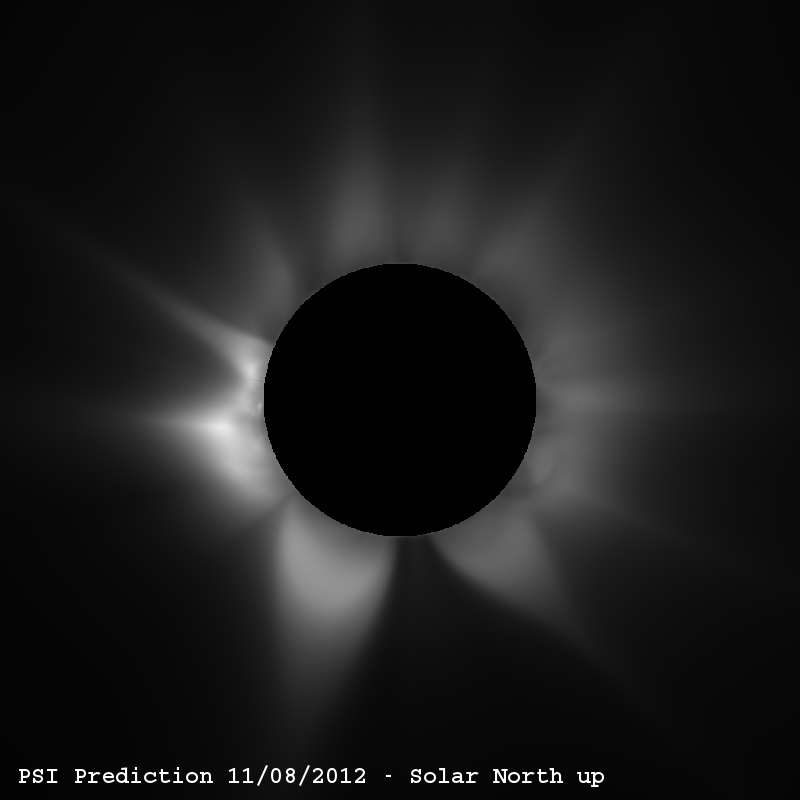}
\end{center}
 \textbf{\refstepcounter{figure}\label{WL_B_CarrRt2130} Figure \arabic{figure}.}{ Predictions of the coronal magnetic field configuration (left) and white-light synthetic image (right) for the solar eclipse of November 13, 2012. The images shown below are aligned such that solar north is vertically upward and includes the tilt due to the solar $B_0$ angle. The date on the panels is when the prediction was made. PSI have been providing predictions for solar eclipses for many years, which can be found at: http://www.predsci.com/corona/.}
\end{figure*}

To develop useful model solutions for the solar corona, we use the magnetohydrodynamic (MHD) approximation, which is appropriate for long-scale, low-frequency phenomena in magnetized plasmas. In the past, we employed a ``\emph{polytropic approximation}" for treating the heating of the coronal plasma and the acceleration of the solar wind \citep{2001JGR...10615889R,2012SoPh..277..355R}. While this approach produces remarkably good solutions for the structure of the coronal magnetic field, this is at the expense of poorer velocity and density profiles. In this study, however, we use our ``\emph{thermodynamic}" model, which relies on coronal heating functions that are guided by observational constraints \citep{2009ApJ...690..902L,2015ApJ...802..105R}. Detailed comparisons with EUV and X-ray observations from several spacecraft have allowed us to constrain the likely functional forms for this heating, such that they reproduce the observed emission.

Most (if not all) of the Hanle effect studies in the literature were based on well-defined magnetic field configurations (e.g., theoretical models or extrapolated photospheric magnetic fields), which were lacking well-defined plasma parameters such as densities, temperatures, and velocities. These quantities enter directly into the definition of the Stokes parameters encompassing the magnetic field signature that is the Hanle effect. Ad-hoc approximations may provide valuable results and order of magnitude estimates of different quantities as well as estimates of linear polarization of a given spectral line, but they cannot yield physically meaningful estimates of the coronal magnetic field through the Hanle effect. These assumptions can be improved upon by considering self-consistent magnetic fields and plasma parameters obtained through MHD simulations.

To obtain realistic estimates of the polarization parameters of the UV \ion{H}{i}~Ly-$\alpha$ (1216~{\AA}), we utilize high-resolution MHD simulations using Predictive Science's state-of-the-art MAS code. The simulation includes a full thermodynamic description of the plasma. All parameters needed for the calculation of the polarization of the spectral line are obtained in a self-consistent fashion, thus removing any need for heuristic assumptions. All quantities are provided on the nodes of spherical grids whose resolution changes with the heliodistance. Figure~\ref{WL_B_CarrRt2130} displays the magnetic field configuration of the solar corona corresponding to Carrington rotation 2130. The synthetic white-light coronal emission is shown in the right-hand-side panel.

For the coronal Hanle effect of the different spectral lines, we use the magnetic field, density, temperature, and velocity data cube. Line-of-sight (LOS) integration is taken into account. All quantities at any given point on the LOS are obtained by interpolation of the simulation data.

\section{\ion{H}{i} Ly-$\alpha$ Solar Disk Radiation}
\begin{figure*}[!ht]
\begin{center}
\includegraphics[width=0.7\textwidth]{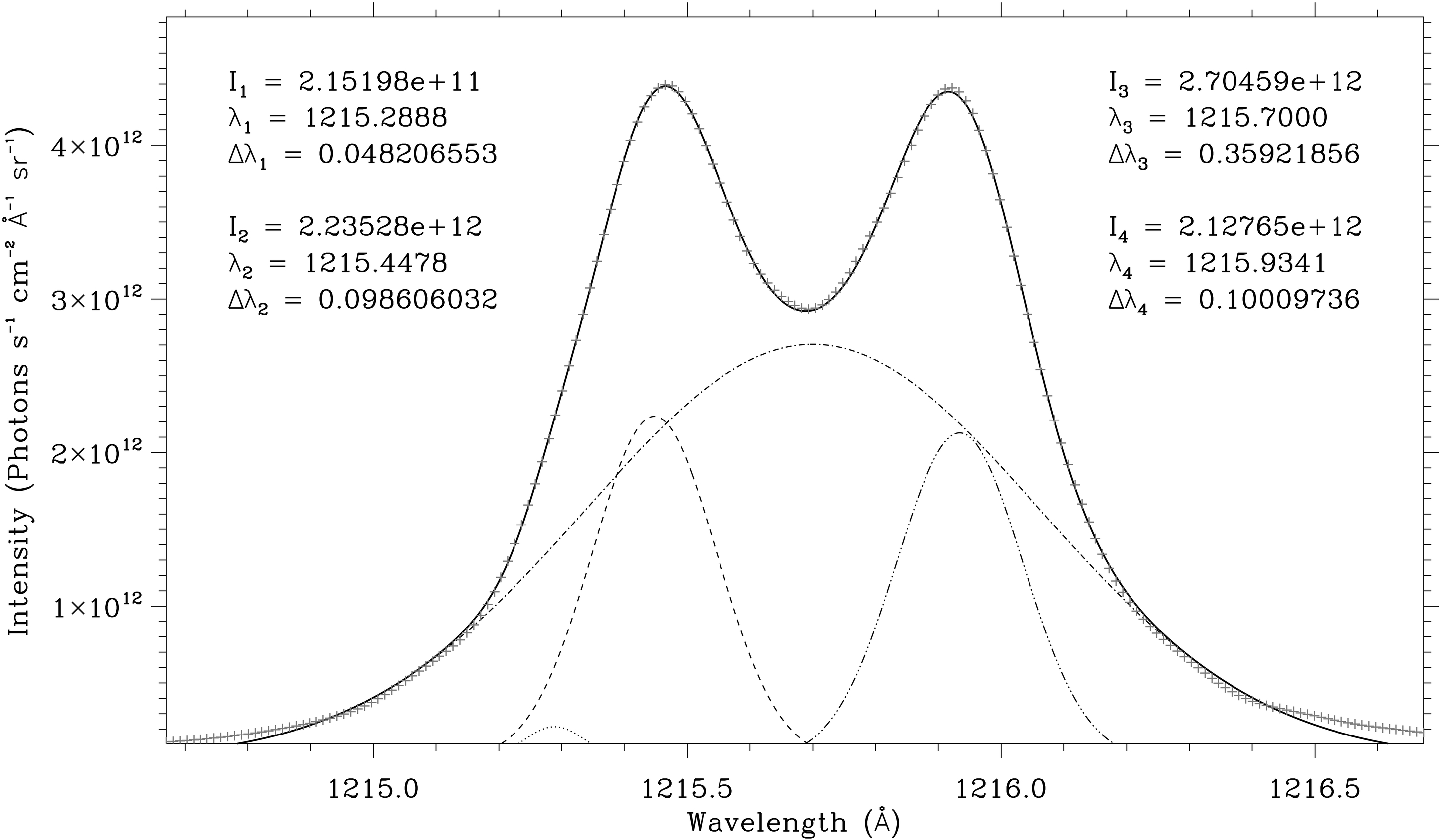}
\end{center}
\textbf{\refstepcounter{figure}\label{H1LyAProfile} Figure \arabic{figure}.}{ Average transition region profile (Solid profile) of the \ion{H}{i} Ly-$\alpha$ profile as observed by SOHO/SUMER \citep{1998A&A...334.1095L}. The "+" signs are the best 4-Gaussian fit of the observed profile. The individual Gaussians are given by the dotted, dashed, dot-dashed, and triple-dot-dashed profiles. The incident \ion{H}{i} Ly-$\alpha$  radiation is assumed to be composed of four individual Gaussian profiles. The parameters of the individual profiles (i.e., amplitude [in photons~cm$^{-2}$~s$^{-1}$~{\AA}$^{-1}$~sr$^{-1}$], central wavelength [in {\AA}] and width [in {\AA}]) are also displayed. }
\end{figure*}

The \ion{H}{i} Ly-$\alpha$ line is formed in the transition region at a temperature $\sim\!1$~MK. The coronal counterpart is formed by scattering of this incident radiation, resulting in the most intense UV coronal line. The solar disk radiation of this line is characterized by very little to no center-to-limb variations \citep{1980ApJ...237L..47B}. The line profile is substantially wider than other lines and is typically complex. Figure~\ref{H1LyAProfile} shows an average \ion{H}{i} Ly-$\alpha$ line profile as observed by SOHO/SUMER \citep{1998A&A...334.1095L}. It is characterized by an inversed peak at the center of the line, making a single Gaussian fit meaningless. For the present study, this profile is fitted with four Gaussians whose parameters are shown in the same figure. Numerically, we assume four individual Gaussian spectral lines (Figure~\ref{H1LyAProfile}) at different frequencies and with different widths and intensities. This assumption is, we believe, the best way to realistically mimic the incident radiation from the solar transition region.

We consider a two level atomic model for the \ion{H}{i} Ly-$\alpha$ line. This assumption is sufficient to describe the light scattering by hydrogen atoms in the solar corona. The spectral line has two components, 2p~$^2$P$^{\rm{o}}_{3/2}\rightarrow$1s~$^2$S$_{1/2}$ (polarizable) and 2p~$^2$P$^{\rm{o}}_{1/2}\rightarrow$1s~$^2$S$_{1/2}$ (non-polarizable), with virtually the same Einstein coefficients of spontaneous emission, $A_{ul}\approx6.2648\times10^8$~s$^{-1}$. The coronal electron collisional component represents less than 1\% of the total intensity \citep{1997SoPh..175..645R}. We neglect this component and assume that the \ion{H}{i} Ly-$\alpha$ coronal line results only from the scattering of incident radiation from the transition region. Since we are interested in the magnetic field, we also neglect the effect of the solar wind velocity. This assumption is justified by the fact that the polarization calculations presented in this paper are achieved at coronal heights lower than 1~$R_\odot$, where the solar wind speed is lower than 100~km~s$^{-1}$. Considering the line width of the incident line, the Doppler distribution effects are neglected. The incident radiation from the solar transition region is unpolarized and the radiation field is assumed to be cylindrically symmetric around the solar vertical, with half-cone angle $\alpha_r$ (i.e., solar disk radiation inhomogeneities [e.g., active regions] are also neglected).

\subsection{The Atomic Model and Polarization}
We consider the case of a two-level atom $(\alpha_lJ_l,\alpha_uJ_u)$ in the presence of a magnetic field $\bf{B}$. We also assume a non-polarizable lower level, such as that of the \ion{H}{i} Ly-$\alpha$ line whose spherically symmetric lower level $^2{\rm{S}}_{1/2}$ (that is not polarizable). Within the frame of the density matrix formalism, the lower level  is described only by its population represented by $^{\alpha_l J_l}\rho^0_0$ within the frame of density matrix formalism. 

We assume that the incident radiation is characterized by a Gaussian spectral profile
\begin{equation}
\displaystyle\mathscr{I}(\Omega,\nu)=\frac{\mathscr{I}_c\ f(\Omega)}{\sqrt{\pi}\ \sigma_i}\ \mathrm{e}^{-\left(\frac{\nu-\nu_0}{\sigma_i}\right)^2},
\end{equation}
where $\mathscr{I}_c$ is the solar disk center radiance (in erg~cm$^{-2}$~s$^{-1}$~sr$^{-1}$~Hz$^{-1}$), $\nu_0$ and $\sigma_i$ are the line center frequency and width of the incident profile, and $f(\Omega)$ describes the center-to-limb variation of the incoming radiation field. In the case of Ly-$\alpha$, $f(\Omega)\approx1$ \citep[see][]{1980ApJ...237L..47B}. In the solar frame, the properties of the incident radiation field are given by
\begin{equation}
\begin{array}{l}
\displaystyle\mathfrak{J}^0_0(\nu)=\oint\frac{\mathrm{d}\Omega}{4\pi}\ \mathscr{I}(\Omega,\nu) \\ 
\displaystyle\mathfrak{J}^2_0(\nu)=\oint\frac{\mathrm{d}\Omega}{4\pi}\ \frac{1}{2\sqrt{2}}(3\cos^2\alpha_r-1)\ \mathscr{I}(\Omega,\nu) 
\end{array}
\end{equation}
All other multipoles (i.e., $\mathfrak{J}^1_{0,\pm1}$ and $\mathfrak{J}^2_{\pm1,\pm2}$) are zero since the incident radiation is not polarized. The density matrix multipoles of the incident radiation have to be re-written in the magnetic field reference frame, which is obtained from the solar frame by a rotation $\mathcal{R}(\psi,\eta,0)$, where $\psi$ and $\eta$ are, respectively, the azimuth and co-latitude of the vector magnetic field with respect to the solar vertical reference frame.

For a two-level atom with unpolarized lower level, the atomic polarization properties of the upper level are given by the atomic density matrix:
\begin{eqnarray}
^{\alpha_uJ_u}\rho^K_Q(\nu,\Omega) &=& \frac{^{\alpha_lJ_l}\rho^0_0(\nu)}{A_{ul}+{\rm{i}}\ Q\ \omega_L}\sqrt{3(2J_l+1)}\ B_{lu}\ (-1)^{1+J_l+J_u+Q}\ 
\left\{ \begin{array}{ccc}
  1 & 1 & K \\
  J_u & J_u & J_l
 \end{array} \right\}
\ ^B\mathfrak{J}^K_{-Q}(\nu,\Omega) \nonumber \\
 &=& \sqrt{\frac{2J_l+1}{2J_u+1}}\ \frac{B_{lu}\ ^{\alpha_lJ_l}\rho^0_0(\nu)}{A_{ul}+{\rm{i}}\ Q\ \omega_L}\ w^{(K)}_{ul}\ (-1)^{Q}\ \ ^B\mathfrak{J}^K_{-Q}(\nu,\Omega)
\label{Hanle2level}
\end{eqnarray}
where $K=0,\cdots,2J_u$ and $Q=-K,\cdots,K$. $\omega_L=2\pi\  g_{u}\ \nu_L$ is the Larmor angular frequency. For details on the derivation of Eq.(\ref{Hanle2level}), see \citet{2004ASSL..307.....L}. $\nu_L$ is the Larmor frequency and reflects the Hanle effect due to the presence of the magnetic field, $g_u$ is the Land\'e factor of the upper level, and $A_{ul}$ and $B_{lu}$ are the Einstein coefficients for the spontaneous emission and absorption from the lower to upper levels. The symbol between the brackets is the Wigner 6j-symbol. Eq.(\ref{Hanle2level}) is the solution of the statistical system of linear equations of the atomic system in the steady-state case.

\section{Hanle effect of the \ion{H}{i} L\MakeLowercase{y}-$\alpha$ coronal lines}

\begin{figure*}[!ht]
\begin{center}
\includegraphics[width=\textwidth]{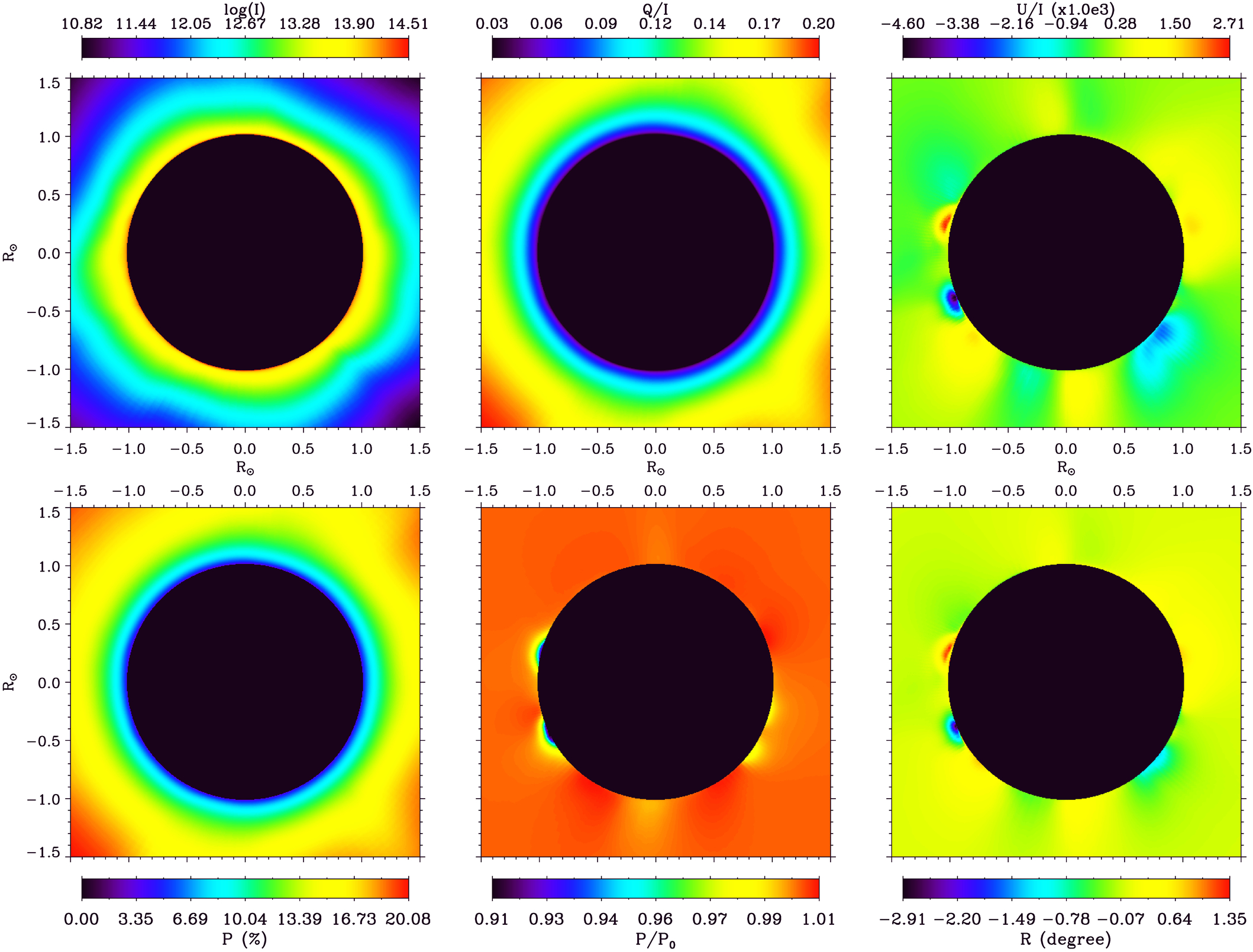}
\label{Ly_Alpha_2xB_new20150310}
\end{center}
\textbf{\refstepcounter{figure}\label{Ly_Alpha_2xB_new20150310} Figure \arabic{figure}.}{ Line-profile-integrated Stokes parameters (top) and linear polarization (bottom) of the coronal \ion{H}{i}~Ly-$\alpha$ line. $P$, $P_0$, and $R$ are the polarization degree, the polarization degree in zero magnetic field, and the rotation of the plane of polarization with respect to the local solar limb, respectively. Above regions of relatively strong coronal magnetic fields the Hanle depolarization attains about 10\% and the rotation of the plane of polarization is about $3^\circ$. This illustrates that although this line is not the best in terms of sensitivity to the Hanle effect, the effects of the magnetic field on the linear polarization are significant. 
 }
\end{figure*}

The results of the forward modeling of the linear polarization of the coronal \ion{H}{i}~Ly-$\alpha$ line are shown in Figure~\ref{Ly_Alpha_2xB_new20150310}. In the absence of the effect of the coronal magnetic field (i.e., Hanle effect), the direction of polarization is parallel to the local solar limb regardless of coronal altitude. The fractional linear polarization increases as a function of altitude because of the increased anisotropy of the incident solar disk radiation. The LOS integration is also more important with increasing altitude above the solar limb. This is due to the increasingly shallower density gradient. For the present calculations, the LOS integration is done for a range of $5~R_\odot$ centered on the plane of the sky.

The top panels of Figure~\ref{Ly_Alpha_2xB_new20150310} illustrated the altitude variation of the Stokes parameters ($\log_{10}I$, $Q/I$, and $U/I$, respectively). The lowest coronal altitude of the calculations is $1.015~R_\odot$. The bottom panels show the degree of linear polarization (in \%, left), depolarization (i.e., the ratio of the degree of polarization to that  in the absence of magnetic field, middle), and the rotation of the plane of linear polarization with respect to the local solar limb (in degrees). It is clear that the signature of the Hanle effect, which is given primarily by the depolarization and the rotation of plane of polarization, is limited to low-altitude regions where the magnetic field is relatively strong. This is expected because of the sensitivity of this line to the Hanle effect.

Although these results are still preliminary, they suggest that the \ion{H}{i}~Ly-$\alpha$ line could be very useful for measuring the coronal magnetic field, particularly at low latitude, and in strong field regions (i.e., above active regions). Theoretical polarization rates (not shown here) in these coronal line are reasonably high and could be easily measured. The depolarization with respect to the zero magnetic field case attains $\sim\!10\%$ in some areas. The main parameter that limits the measurability of the Hanle effect in this line is the rotation of the plane of polarization. Above active regions, a rotation of about $3^\circ$ is obtained, and rotations of more than $1^\circ$ are obtained in larger areas. These results show that this line is promising in terms of constraining the coronal magnetic field, despite the fact that it is not the most suitable line in terms of sensitivity to the coronal Hanle effect. In combination with other spectral lines with complementary sensitivities to the effect of the coronal magnetic field, the Hanle effect could provide reliable constraints on the coronal magnetic field.

\section{Hanle effect of the \ion{H\MakeLowercase{e}}{i} 10830~{\AA} coronal line}

\begin{figure*}[!ht]
\begin{center}
\includegraphics[width=0.95\textwidth]{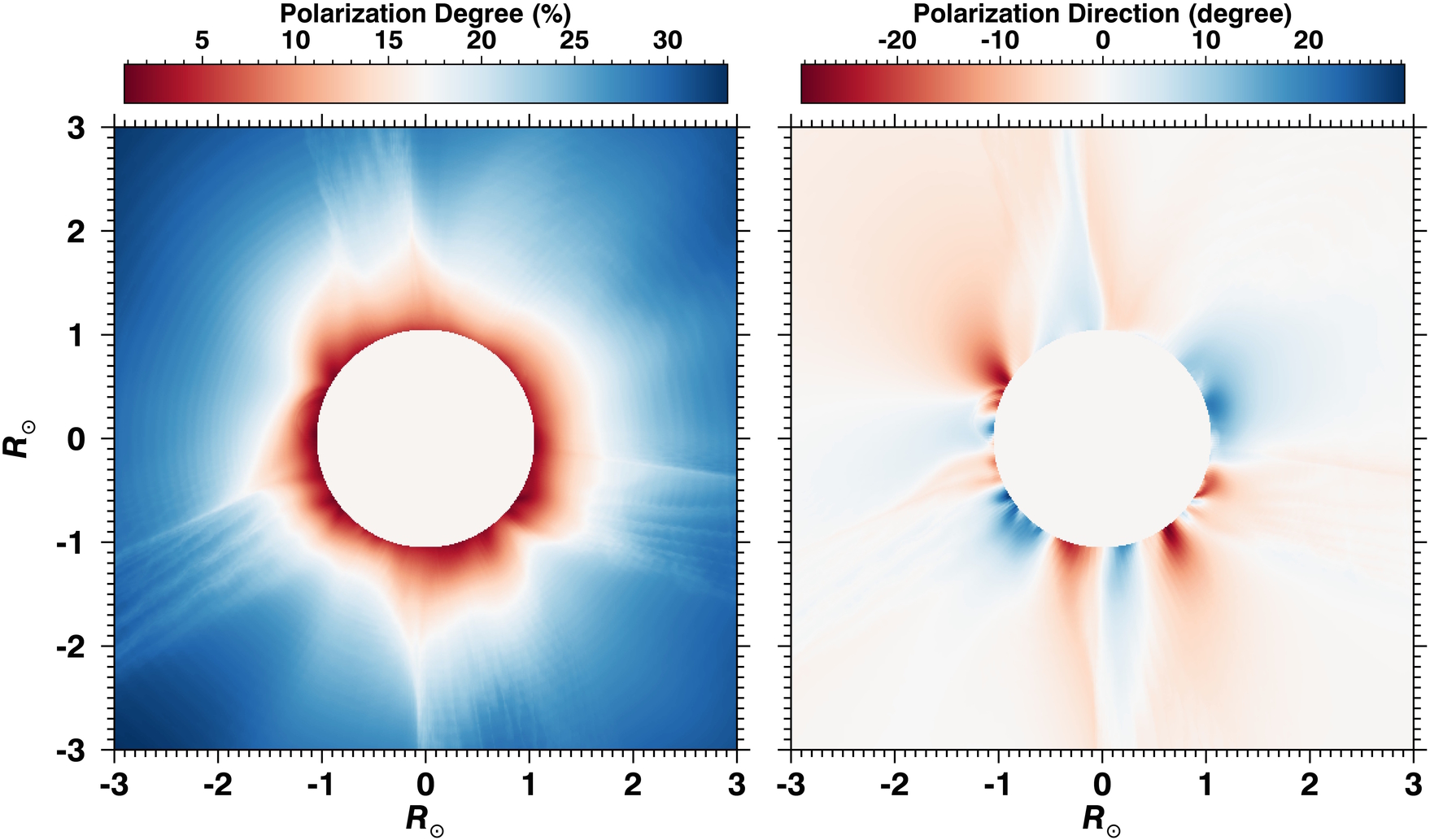}
\end{center}
\textbf{\refstepcounter{figure}\label{He_10830_Hanle_main_LOSintg_2} Figure \arabic{figure}.}{ Polarization parameters of the He I 10830 line. (Left) polarization degree (in \%) and (Right) rotation of the plane of polarization (in degrees) with respect to the local tangent to the solar limb. }
\end{figure*}

\citet{2007ApJ...667L.203K} showed evidence for an extended diffuse surface brightness flux at the \ion{He}{i}~10830~{\AA} line using observations from the SOLARC coronagraph. The observations show that emissions result from cold helium (i.e., narrow line profiles), which is unlikely to be scattered by the solar wind helium that is presumably significantly hotter. The authors argue that cold helium atoms form on dust grains, which provide an atomic population different to that of the solar wind. The importance of these observations stems from the sensitivity of the \ion{He}{i}~10830~{\AA} line to the Hanle effect, which corresponds to magnetic field strength ranging from $\sim\!0.2$~Gauss to $<\!10$~Gauss. This line may be the most suitable line for the diagnostic of coronal magnetic fields through the Hanle effect.

Figure~\ref{He_10830_Hanle_main_LOSintg_2} shows forward modeling results of the linear polarization of the \ion{He}{i} 10830~{\AA}. The LOS-integration scheme is the same as for \ion{H}{i}~Ly-$\alpha$. Unlike \ion{H}{i}~Ly-$\alpha$ where the Hanle effect is limited to low-height, strong field regions, the linear polarization of  \ion{He}{i} 10830~{\AA} shows more variations as it depicts coronal structures, such as streamers, closed field regions. The variation of the polarization parameters spreads over larger intervals, which make their measurement easier. This is expected because of the higher sensitivity of this line to the relatively weak coronal magnetic fields. \citet{10.3389/fspas.2016.00013} present a complementary analysis of the polarization of the \ion{He}{i} 10830~{\AA} line.

We believe that the linear polarization of the \ion{He}{i} 10830~{\AA} line could provide valuable constraints on the coronal magnetic field. In combination with UV lines such as \ion{H}{i}~Ly-$\alpha$, -$\beta$, and  \ion{O}{vi}~103.2~{nm}, as well as IR forbidden lines, the coronal Hanle effect could provide reliable diagnostic of the coronal magnetic field and consequently to extrapolation and MHD models.

\section{Summary and Conclusions}

The preliminary results of the forward modeling of the linear polarization of two coronal spectral lines (i.e., \ion{H}{i}~Ly-$\alpha$ and \ion{He}{i} 10830~{\AA}) with different sensitivities to the Hanle effect are promising. The UV \ion{H}{i}~Ly-$\alpha$ line is mainly sensitive to low-height, strong magnetic fields above active regions. The polarization of the \ion{He}{i} 10830~{\AA} line, which is sensitive to magnetic fields ranging roughly from $\sim\!0.2$ Gauss up to $<\!10$~Gauss, shows more variations with coronal height and traces different coronal structures with different magnetic topologies (e.g., streamers and closed field regions). 

Forward modeling of the Hanle effect is an important step in our quest for direct measurements of the magnetic field in the solar corona, which is a very difficult problem that includes different issues that observations will be subject to (e.g., $180^\circ$ and Van Vleck $90^\circ$ ambiguities and LOS-integration). Forward modeling will allow us to fully understand these problems and develop the necessary tools to analyze the observations. It also shows the potential of the Hanle effect in different wavelength regimes, which can be utilized for future space solar missions with UV and IR polarimeters  \citep[e.g.,][]{2012ExA....33..271P}.

We believe that the combination of the linear polarization of coronal lines with complementary sensitivities to the Hanle effect is promising and could provide long-sought measurements of the coronal magnetic field. The Hanle effect is a powerful tool that may provide the most reliable diagnostics of the magnetic field at relatively low coronal heights. Radio observations along with polarimetric measurements in the Zeeman and saturated Hanle regimes may provide complementary constrains that could help help piece together the coronal magnetic field structure.

\section*{Acknowledgments}
We, the authors, would like to thank the anonymous reviewers for the constructive comments and criticism that helped improving the quality of the manuscript. This work was enabled by discussions with members of the International Space Science Institute (ISSI) working group on coronal magnetism (2013-2014), particularly J. Kuhn. SG acknowledges support from the Air Force Office of Space Research, FA9550-15-1-0030; NCAR is supported by the National Science Foundation.

\bibliographystyle{frontiersinSCNS} 	

\begin{thebibliography}{54}
\providecommand{\natexlab}[1]{#1}
\expandafter\ifx\csname urlstyle\endcsname\relax
  \providecommand{\doi}[1]{doi:\discretionary{}{}{}#1}\else
  \providecommand{\doi}{doi:\discretionary{}{}{}\begingroup
  \urlstyle{rm}\Url}\fi
\providecommand{\selectlanguage}[1]{\relax}

\bibitem[{\textbf{{Arnaud}}(1982a)}]{1982A&A...112..350A}
{Arnaud}, J., {Observed polarization of the Fe XIV 5303 coronal emission line}, \emph{\aap}, 112, 350 (1982a).

\bibitem[{\textbf{{Arnaud}}(1982b)}]{1982A&A...116..248A}
{Arnaud}, J., {The analysis of \ion{Fe}{xiv} 5303 coronal emission-line polarization measurements}, \emph{\aap}, 116, 248 (1982b).

\bibitem[{\textbf{{Arnaud} \& {Newkirk}}(1987)}]{1987A&A...178..263A}
{Arnaud}, J. and {Newkirk}, G., Jr., {Mean properties of the polarization of the \ion{Fe}{xiii} 10747~{\AA} coronal emission line}, \emph{\aap}, 178, 263 (1987).

\bibitem[{\textbf{{Bommier} \& {Sahal-Br\'echot}}(1978)}]{1978A&A....69...57B}
{Bommier}, V. and {Sahal-Br\'echot}, S., {Quantum theory of the Hanle
  effect - Calculations of the Stokes parameters of the D$_3$ helium line for
  quiescent prominences}, \emph{\aap}, 69, 57 (1978).

\bibitem[{\textbf{{Bommier}}(1980)}]{1980A&A....87..109B}
{Bommier}, V., {Quantum theory of the Hanle effect. II - Effect of
  level-crossings and anti-level-crossings on the polarization of the D$_3$ helium
  line of solar prominences}, \emph{\aap}, 87, 109 (1980).

\bibitem[{\textbf{{Bommier} et~al.}(1981)\textbf{{Bommier}, {Sahal-Br\'echot},
  and {Leroy}}}]{1981A&A...100..231B}
{Bommier}, V., {Sahal-Br\'echot}, S., and {Leroy}, J.~L., {Determination
  of the complete vector magnetic field in solar prominences, using the Hanle
  effect}, \emph{\aap}, 100, 231 (1981).
  
\bibitem[{\textbf{{Bommier} \& {Sahal-Br\'echot}}(1982)}]{1982SoPh...78..157B}
{Bommier}, V. and {Sahal-Br\'echot}, S., {The Hanle effect of the coronal L-$\alpha$ line of hydrogen - Theoretical investigation}, \emph{\solphys}, 78, 157 (1982). doi:10.1007/BF00151151

\bibitem[{\textbf{{Bommier} et~al.}(1994)\textbf{{Bommier}, {Landi
  Degl'Innocenti}, {Leroy}, and {Sahal-Br\'echot}}}]{1994SoPh..154..231B}
{Bommier}, V., {Landi Degl'Innocenti}, E., {Leroy}, J.-L., and {Sahal-Br\'echot},
  S., {Complete determination of the magnetic field vector and of the
  electron density in 14 prominences from linear polarization measurements in
  the \ion{He}{i}~D$_3$ and H-$\alpha$ lines}, \emph{\solphys}, 154, 231 (1994).
  \doi{10.1007/BF00681098}

\bibitem[{\textbf{{Bonnet} et~al.}(1980)\textbf{{Bonnet}, {Decaudin}, {Bruner},
  {Acton}, and {Brown}}}]{1980ApJ...237L..47B}
{Bonnet}, R.~M., {Decaudin}, M., {Bruner}, E.~C., Jr., {Acton}, L.~W., and
  {Brown}, W.~A., {High-resolution Lyman-$\alpha$ filtergrams of the Sun},
  \emph{\apjl}, 237, L47 (1980). \doi{10.1086/183232}

\bibitem[{\textbf{{Casini} \& {Judge}}(1999)}]{1999ApJ...522..524C}
{Casini}, R. and {Judge}, P.~G., {Spectral Lines for Polarization
  Measurements of the Coronal Magnetic Field. II. Consistent Treatment of the
  Stokes Vector for Magnetic-Dipole Transitions}, \emph{\apj}, 522, 524 (1999).
  \doi{10.1086/307629}

\bibitem[{\textbf{{Charvin}}(1965)}]{1965AnAp...28..877C}
{Charvin}, P., {{\'E}tude de la polarisation des raies interdites de la
  couronne solaire. Application au cas de la raie verte {$\lambda$}5303},
  \emph{Annales d'Astrophysique}, 28, 877 (1965).

\bibitem[{\textbf{{Dima} et~al.}(2016)\textbf{{Dima}, {Kuhn}, and
  {Berdyugina}}}]{10.3389/fspas.2016.00013}
{Dima}, G.~I., {Kuhn}, J.~R., and {Berdyugina}, S.~V., {Infrared
  Dual-line Hanle diagnostic of the Coronal Vector Magnetic Field},
  \emph{\frass}, 3, 13 (2016). \doi{10.3389/fspas.2016.00013}

\bibitem[{\textbf{{Domingo} et~al.}(1995)\textbf{{Domingo}, {Fleck}, and {Poland}}}]{1995SoPh..162....1D}
{Domingo}, V., {Fleck}, B., and {Poland}, A.~I., {The SOHO Mission: an Overview},
  \emph{\solphys}, 162, 1 (1995). \doi{10.1007/BF00733425}

\bibitem[{\textbf{{Fineschi} et~al.}(1991)\textbf{{Fineschi}, {Hoover},
  {Fontenla}, and {Walker}}}]{1991OptEn..30.1161F}
{Fineschi}, S., {Hoover}, R.~B., {Fontenla}, J.~M., and {Walker}, A.~B.~C., Jr., {Polarimetry of extreme ultraviolet lines in solar astronomy}, \emph{Optical Engineering}, 30, 1161 (1991). \doi{10.1117/12.55922}

\bibitem[{\textbf{{Fineschi} et~al.}(1993)\textbf{{Fineschi}, {Chiuderi},
  {Poletto}, {Hoover}, and {Walker}}}]{1993MmSAI..64..441F}
{Fineschi}, S., {Chiuderi}, C., {Poletto}, G., {Hoover}, R.~B., and {Walker},
  A.~B.~C., Jr., {LY-A-CO-PO (LY-{$\alpha$} coronograph/polarimeter): an
  instrument to measure coronal magnetic fields}, \emph{\memsai}, 64, 441 (1993).

\bibitem[{\textbf{{Fineschi} \& {Habbal}}(1995)}]{1995sowi.confR..68F}
{Fineschi}, S. and {Habbal}, S.~R., {Coronal magnetic field diagnostics
  via the Hanle effect of Lyman series lines}, Proc. of Solar Wind 8, 68 (1995).

\bibitem[{\textbf{{Fineschi} et~al.}(1999)\textbf{{Fineschi}, {Gardner},
  {Kohl} et~al.}}]{1999SPIE.3764..147F}
{Fineschi}, S., {Gardner}, L.~D., {Kohl}, et~al., {Polarimetry of the UV solar corona with ASCE},
  in Ultraviolet and X-Ray Detection, Spectroscopy, and Polarimetry III, \emph{\procspie}, vol. 3764, 147 (1999).

\bibitem[{\textbf{{Fineschi}}(2001)}]{2001ASPC..248..597F}
{Fineschi}, S., {Space-based Instrumentation for Magnetic Field Studies
  of Solar and Stellar Atmospheres}, in Magnetic Fields Across the Hertzsprung-Russell
  Diagram, \emph{\aspconf}, vol. 248, 597 (2001).

\bibitem[{\textbf{{Gibson} et~al.}(2016)\textbf{{Gibson}, {Kucera}, {White} et~al.}}]{10.3389/fspas.2016.00008}
{Gibson}, S., {Kucera}, T., {White}, S.~M., et~al., {FORWARD: A toolset for multiwavelength coronal magnetometry}, \emph{\frass}, 3, 8 (2016). \doi{10.3389/fspas.2016.00008}

\bibitem[{\textbf{{Habbal} et~al.}(2001)\textbf{{Habbal}, {Woo}, and {Arnaud}}}]{2001ApJ...558..852H}
{Habbal}, S.~R., {Woo}, R., and {Arnaud}, J., {On the Predominance of
  the Radial Component of the Magnetic Field in the Solar Corona}, \emph{\apj},
  558, 852 (2001). \doi{10.1086/322308}

\bibitem[{\textbf{{Hanle}}(1924)}]{1924ZPhy...30...93H}
{Hanle}, W., {{\"U}ber magnetische Beeinflussung der Polarisation der
  Resonanzfluoreszenz}, \emph{Zeitschrift f\"ur Physik}, 30, 93 (1924). \doi{10.1007/BF01331827}

\bibitem[{\textbf{{Hassler} et~al.}(1997)\textbf{{Hassler}, {Lemaire}, {Longval}}}]{1997ApOpt..36..353H}
{Hassler}, D.~M., {Lemaire}, P., {Longval}, Y., {Polarization sensitivity of the SUMER instrument on SOHO}, \emph{\ao}, 36, 353 (1997). \doi{10.1364/AO.36.000353}

\bibitem[{\textbf{{Kohl} et~al.}(1995)\textbf{{Kohl}, {Esser}, {Gardner} et~al.}}]{1995SoPh..162..313K}
{Kohl}, J.~L., {Esser}, R., {Gardner}, L.~D., et~al., {The Ultraviolet Coronagraph Spectrometer for the Solar and Heliospheric Observatory}, \emph{\solphys}, 162, 313 (1995). \doi{10.1007/BF00733433}

\bibitem[{\textbf{{Kohl} et~al.}(1998)\textbf{{Kohl}, {Noci}, {Antonucci} et~al.}}]{1998ApJ...501L.127K}
{Kohl}, J.~L., {Noci}, G., {Antonucci}, E., et~al., {UVCS/SOHO Empirical Determinations of
  Anisotropic Velocity Distributions in the Solar Corona}, \emph{\apjl}, 501, L127 (1998). \doi{10.1086/311434}

\bibitem[{\textbf{{Kuckein} et~al.}(2009)\textbf{{Kuckein}, {Centeno}, {Mart{\'{\i}}nez Pillet} et~al.}}]{2009A&A...501.1113K}
{Kuckein}, C., {Centeno}, R., {Mart{\'{\i}}nez Pillet}, V., et~al., {Magnetic field strength of active region filaments}, \emph{\aap}, 501, 1113 (2009). \doi{10.1051/0004-6361/200911800}

\bibitem[{\textbf{{Kuckein} et~al.}(2012)\textbf{{Kuckein}, {Mart{\'{\i}}nez Pillet}, {Centeno}}}]{2012A&A...539A.131K}
{Kuckein}, C., {Mart{\'{\i}}nez Pillet}, V., and {Centeno}, R., {An active region filament studied simultaneously in the chromosphere and photosphere. I. Magnetic structure}, \emph{\aap}, 539, A131 (2012). \doi{10.1051/0004-6361/201117675}

\bibitem[{\textbf{{Kuhn} et~al.}(2007)\textbf{{Kuhn}, {Arnaud}, {Jaeggli}, {Lin}, and {Moise}}}]{2007ApJ...667L.203K}
{Kuhn}, J.~R., {Arnaud}, J., {Jaeggli}, S., {Lin}, H., and {Moise}, E.,
  {Detection of an Extended Near-Sun Neutral Helium Cloud from Ground-based
  Infrared Coronagraph Spectropolarimetry}, \emph{\apjl}, 667, L203 (2007). \doi{10.1086/522370}

\bibitem[{\textbf{{Lagg} et~al.}(2004)\textbf{{Lagg}, {Woch}, {Krupp}, and {Solanki}}}]{2004A&A...414.1109L}
{Lagg}, A., {Woch}, J., {Krupp}, N., {Solanki}, S.~K., {Retrieval of the full magnetic vector with the \ion{He}{i} multiplet at 1083 nm. Maps of an emerging flux region}, \emph{\aap}, 414, 1109 (2004). \doi{10.1051/0004-6361:20031643}

\bibitem[{\textbf{{Landi Degl'Innocenti}}(1982)}]{1982SoPh...79..291L}
{Landi Degl'Innocenti}, E., {The determination of vector magnetic fields
  in prominences from the observations of the Stokes profiles in the D$_3$ line of
  helium}, \emph{\solphys}, 79, 291 (1982). \doi{10.1007/BF00146246}

\bibitem[{\textbf{{Landi Degl'Innocenti} and {Landolfi}}(2004)}]{2004ASSL..307.....L}
{Landi Degl'Innocenti}, E. and {Landolfi}, M., {Polarization in Spectral Lines}, \emph{Astrophysics and Space Science Library}, vol. 307 (2004). \doi{10.1007/978-1-4020-2415-3}

\bibitem[{\textbf{{Lemaire} et~al.}(1998)\textbf{{Lemaire}, {Emerich}, {Curdt},
  {Schuehle}, and {Wilhelm}}}]{1998A&A...334.1095L}
{Lemaire}, P., {Emerich}, C., {Curdt}, W., {Schuehle}, U., and {Wilhelm}, K., {Solar \ion{H}{i} Lyman-$\alpha$ full disk profile obtained with the SUMER/SOHO spectrometer}, \emph{\aap}, 334, 1095 (1998).

\bibitem[{\textbf{{Leroy} et~al.}(1977)\textbf{{Leroy}, {Ratier}, and
  {Bommier}}}]{1977A&A....54..811L}
{Leroy}, J.~L., {Ratier}, G., and {Bommier}, V., {The polarization of the D$_3$ emission line in prominences}, \emph{\aap}, 54, 811 (1977).

\bibitem[{\textbf{{Leroy} et~al.}(1983)\textbf{{Leroy}, {Bommier}, and
  {Sahal-Br\'echot}}}]{1983SoPh...83..135L}
{Leroy}, J.~L., {Bommier}, V., and {Sahal-Br\'echot}, S., {Solar Cycles, Solar Magnetic Field, Solar Prominences, Angular Resolution, Magnetic Field Configurations, Periodic Variations, Photosphere, Polar Regions}, \emph{\solphys}, 83, 135 (1983). \doi{10.1007/BF00148248}

\bibitem[{\textbf{{Leroy} et~al.}(1984)\textbf{{Leroy}, {Bommier}, and {Sahal-Br\'echot}}}]{1984A&A...131...33L}
{Leroy}, J.~L., {Bommier}, V., and {Sahal-Br\'echot}, S., {New data on the magnetic structure of quiescent prominences}, \emph{\aap}, 131, 33 (1984).

\bibitem[{\textbf{{Li} et~al.}(1998)\textbf{{Li}, {Habbal}, {Kohl}, and
  {Noci}}}]{1998ApJ...501L.133L}
{Li}, X., {Habbal}, S.~R., {Kohl}, J.~L., and {Noci}, G., {The Effect of
  Temperature Anisotropy on Observations of Doppler Dimming and Pumping in the
  Inner Corona}, \emph{\apjl}, 501, L133 (1998). \doi{10.1086/311428}

\bibitem[{\textbf{{Lionello} et~al.}(2009)\textbf{{Lionello}, {Linker}, and {Miki\'c}}}]{2009ApJ...690..902L}
{Lionello}, R., {Linker}, J.~A., and {Miki\'c}, Z., {Multispectral emission of the Sun during the first
whole Sun month: Magnetohydrodynamic simulations}, \emph{\apj}, 690, 902 (2009). \doi{10.1088/0004-637X/690/1/902}

\bibitem[{\textbf{{Lin} et~al.}(2004)\textbf{{Lin}, {Kuhn}, and
  {Coulter}}}]{2004ApJ...613L.177L}
{Lin}, H., {Kuhn}, J.~R., and {Coulter}, R., {Coronal Magnetic Field
  Measurements}, \emph{\apjl}, 613, L177 (2004). \doi{10.1086/425217}

\bibitem[{\textbf{{L{\'o}pez Ariste} \& {Casini}}(2002)}]{2002ApJ...575..529L}
{L{\'o}pez Ariste}, A. and {Casini}, R., {Magnetic Fields in
  Prominences: Inversion Techniques for Spectropolarimetric Data of the \ion{He}{i}
  D$_{3}$ Line}, \emph{\apj}, 575, 529 (2002). \doi{10.1086/341260}

\bibitem[{\textbf{{Manso Sainz} \& {Trujillo Bueno}}(2009)}]{2009ASPC..405..423M}
{Manso Sainz}, R. and {Trujillo Bueno}, J., {A Possible Polarization Mechanism of EUV Coronal Lines}, in Solar Polarization 5, \emph{\aspconf}, vol. 405, 423 (2009).

\bibitem[{\textbf{{Merenda} et~al.}(2011)\textbf{{Merenda}, {Lagg}, and
  {Solanki}}}]{2011A&A...532A..63M}
{Merenda}, L., {Lagg}, A., and {Solanki}, S.~K., {The height of chromospheric loops in an emerging flux region}, \emph{\aap}, 532, A63 (2011). \doi{10.1051/0004-6361/201014988}

\bibitem[{\textbf{{Peter} et~al.}(2012)\textbf{{Peter}, {Abbo}, {Andretta} et~al.}}]{2012ExA....33..271P}
{Peter}, H., {Abbo}, L., {Andretta}, E.~A., et~al., {Solar magnetism eXplorer (SolmeX).
  Exploring the magnetic field in the upper atmosphere of our closest star},
  \emph{Experimental Astronomy}, 33, 271 (2012). \doi{10.1007/s10686-011-9271-0}

\bibitem[{\textbf{{Querfeld}}(1974)}]{1974psns.coll..254Q}
{Querfeld}, C.~W., {The High Altitude Observatory Coronal-Emmission Polarimeter}, in Planets, Stars, and
  Nebulae: Studied with Photopolarimetry, IAU Colloq. 23, 254 (1974).

\bibitem[{\textbf{{Querfeld} \& {Elmore}}(1976)}]{1976BAAS....8..368Q}
{Querfeld}, C.~W. and {Elmore}, D.~E., {Observation of Polarization in \ion{Fe}{xiii} 10747 {\AA} Coronal Emission Line}, \baas, vol. 8, 368 (1976).

\bibitem[{\textbf{{Querfeld}}(1977)}]{1977ROLun..12..109Q}
{Querfeld}, C.~W., {Observations of \ion{Fe}{xiii} 10747 {\AA} coronal emission-line polarization.}, \emph{Reports of the Lund Observatory}, 12, 109 (1977).

\bibitem[{\textbf{{Querfeld} et~al.}(1985)\textbf{{Querfeld}, {Smartt}, {Bommier}, {Landi Degl'Innocenti}, and {House}}}]{1985SoPh...96..277Q}
{Querfeld}, C.~W., {Smartt}, R.~N., {Bommier}, V., {Landi Degl'Innocenti}, E., and {House}, L.~L., {Vector magnetic fields in prominences. II - He I D$_3$ Stokes profiles analysis for two quiescent prominences}, \emph{\solphys}, 96, 277 (1985). \doi{10.1007/BF00149684}

\bibitem[{\textbf{{Raouafi} et~al.}(1999{\natexlab{a}})\textbf{{Raouafi},
  {Lemaire}, and {Sahal-Br{\'e}chot}}}]{1999A&A...345..999R}
{Raouafi}, N.-E., {Lemaire}, P., and {Sahal-Br{\'e}chot}, S., {Detection of the \ion{O}{vi} 103.2 nm line
  polarization by the SUMER spectrometer on the SOHO spacecraft}, \emph{\aap},
  345, 999 (1999{\natexlab{a}}).

\bibitem[{\textbf{{Raouafi} et~al.}(1999{\natexlab{b}})\textbf{{Raouafi},
  {Sahal-Br{\'e}chot}, {Lemaire}, and {Bommier}}}]{1999ASSL..243..349R}
{Raouafi}, N.-E., {Sahal-Br{\'e}chot}, S., {Lemaire}, P., and {Bommier}, V., {Doppler redistribution of resonance polarization of
  the O VI 103. 2 nm line observed above a polar hole}, Proc. of Solar Polarization 2, \emph{Astrophysics and Space Science Library}, vol. 243, 349 (1999{\natexlab{b}}).

\bibitem[{\textbf{{Raouafi}}(2002)}]{2002A&A...386..721R}
{Raouafi}, N.-E., {Stokes parameters of resonance lines scattered by a moving, magnetic medium. Theory of the two-level atom}, \emph{\aap}, 386, 721 (2002). \doi{10.1051/0004-6361:20020113}

\bibitem[{\textbf{{Raouafi} et~al.}(2002)\textbf{{Raouafi},
  {Sahal-Br{\'e}chot}, and {Lemaire}}}]{2002A&A...396.1019R}
{Raouafi}, N.-E., {Sahal-Br{\'e}chot}, S., and {Lemaire}, P., {Linear
  polarization of the O VI lambda 1031.92 coronal line. II. Constraints on the
  magnetic field and the solar wind velocity field vectors in the coronal polar
  holes}, \emph{\aap}, 396, 1019 (2002). \doi{10.1051/0004-6361:20021418}

\bibitem[{\textbf{{Raouafi}}(2005)}]{2005ESASP.596E...3R}
{Raouafi}, N.-E., {Measurement Methods for Chromospheric and Coronal Magnetic Fields}, in Chromospheric and Coronal Magnetic Fields, \emph{ESA Special Publication}, vol. 596, 3 (2005).

\bibitem[{\textbf{{Raouafi}}(2011)}]{2011ASPC..437...99R}
{Raouafi}, N.-E., {Coronal Polarization}, Proc. of Solar Polarization 6, \emph{\aspconf}, vol. 437, 99 (2011).

\bibitem[{\textbf{{Raymond} et~al.}(1997)\textbf{{Raymond}, {Kohl}, {Noci} et~al.}}]{1997SoPh..175..645R}
{Raymond}, J.~C., {Kohl}, J.~L., {Noci}, G., et~al., {Composition of Coronal Streamers from the
  SOHO Ultraviolet Coronagraph Spectrometer}, \emph{\solphys}, 175, 645 (1997). \doi{10.1023/A:1004948423169}

\bibitem[{\textbf{{Reeves} \& {Parkinson}}(1970)}]{1970ApJS...21....1R}
{Reeves}, E.~M. and {Parkinson}, W.~H., {An Atlas of Extreme-Ultraviolet
  Spectroheliograms from OSO-IV}, \emph{\apjs}, 21, 1 (1970). \doi{10.1086/190217}

\bibitem[{\textbf{{Riley} et~al.}(2001)\textbf{{Riley}, {Linker}, and {Miki\'c}}}]{2001JGR...10615889R}
{Riley}, P., {Linker}, J.~A., and {Miki\'c}, Z., {An empirically-­driven global MHD model of the solar
corona and inner heliosphere}, \emph{\jgr}, 106, 15889 (2001). doi{10.1029/2000JA000121}
  
\bibitem[{\textbf{{Riley} \& Luhmann}(2012)\textbf{{Riley}, and {Luhmann}}}]{2012SoPh..277..355R}
{Riley}, P., and {Luhmann}, J.~G., {Interplanetary signatures of unipolar streamers and the origin of
the slow solar wind}, \emph{\solphys}, 277, 355 (2012). doi{10.1007/s11207-011-9909-0}

\bibitem[{\textbf{{Riley} et~al.}(2012)\textbf{{Riley}, {Linker}, {Lionello}, and {Miki\'c}}}]{2012JASTP..83....1R}
{Riley}, P., {Linker}, J.~A., {Lionello}, R., and {Miki\'c}, Z., {Corotating interaction regions during the recent solar minimum: The power and limitations of global MHD modeling}, \emph{Journal of Atmospheric and Solar-Terrestrial Physics}, 83, 1 (2012). doi{10.1016/j.jastp.2011.12.013}

\bibitem[{\textbf{{Riley} et~al.}(2015)\textbf{{Riley}, {Lionello}, {Linker} et al.}}]{2015ApJ...802..105R}
{Riley}, P., {Lionello}, R., {Linker}, et~al., {Inferring the Structure of the Solar Corona and Inner Heliosphere During the Maunder Minimum Using Global Thermodynamic Magnetohydrodynamic Simulations}, \emph{\apj}, 802, 105 (2015). doi{10.1088/0004-637X/802/2/105}
  
\bibitem[{\textbf{{Sahal-Br\'echot} et~al.}(1977)\textbf{{Sahal-Br\'echot},
  {Bommier}, and {Leroy}}}]{1977A&A....59..223S}
{Sahal-Br\'echot}, S., {Bommier}, V., and {Leroy}, J.~L., {The Hanle
  effect and the determination of magnetic fields in solar prominences},
  \emph{\aap}, 59, 223 (1977).

\bibitem[{\textbf{{Sahal-Br\'echot} et~al.}(1986)\textbf{{Sahal-Br\'echot},
  {Malinovsky}, and {Bommier}}}]{1986A&A...168..284S}
{Sahal-Br\'echot}, S., {Malinovsky}, M., and {Bommier}, V., {The polarization of the \ion{O}{vi} 1032 {\AA} line as a probe for measuring the coronal vector magnetic field via the Hanle effect}, \emph{\aap}, 168, 284 (1986).

\bibitem[{\textbf{{Sasso} et~al.}(2011)\textbf{{Sasso}, {Lagg}, and {Solanki}}}]{2011A&A...526A..42S}
{Sasso}, C., {Lagg}, A., and {Solanki}, S.~K., {Multicomponent \ion{He}{i} 10830 {\AA} profiles in an active filament}, \emph{\aap}, 526, A42 (2011). doi{10.1051/0004-6361/200912956}

\bibitem[{\textbf{{Sasso} et~al.}(2014)\textbf{{Sasso}, {Lagg}, and {Solanki}}}]{2014A&A...561A..98S}
{Sasso}, C., {Lagg}, A., and {Solanki}, S.~K., {Magnetic structure of an activated filament in a flaring active region}, \emph{\aap}, 561, A98 (2014). doi{10.1051/0004-6361/201322481}

\bibitem[{\textbf{{Solanki} et~al.}(2003)\textbf{{Solanki}, {Lagg}, {Woch},
  {Krupp}, and {Collados}}}]{2003Natur.425..692S}
{Solanki}, S.~K., {Lagg}, A., {Woch}, J., {Krupp}, N., and {Collados}, M., {Three-dimensional magnetic field topology in a region of solar coronal heating}, \emph{\nat}, 425, 692 (2003). \doi{10.1038/nature02035}

\bibitem[{\textbf{{Stenflo}}(1982)}]{1982SoPh...80..209S}
{Stenflo}, J.~O., {The Hanle effect and the diagnostics of turbulent
  magnetic fields in the solar atmosphere}, \emph{\solphys}, 80, 209 (1982). \doi{10.1007/BF00147969}

\bibitem[{\textbf{{Tomczyk} et~al.}(2008)\textbf{{Tomczyk}, {Card}, {Darnell} et~al.}}]{2008SoPh..247..411T}
{Tomczyk}, K., {Card}, W., {Darnell}, E., et~al., {An Instrument to Measure Coronal Emission Line Polarization}, \emph{\solphys}, 247, 411 (2008). \doi{10.1007/s11207-007-9103-6}

\bibitem[{\textbf{{Trujillo Bueno} et~al.}(2002{\natexlab{a}})\textbf{{Trujillo Bueno}, {Casini}, {Landolfi}, and {Landi
  Degl'Innocenti}}}]{2002ApJ...566L..53T}
{Trujillo Bueno}, J., {Casini}, R., {Landolfi}, M., and {Landi Degl'Innocenti},
  E., {The Physical Origin of the Scattering Polarization
  of the \ion{Na}{i} D Lines in the Presence of Weak Magnetic Fields}, \emph{\apjl},
  566, L53 (2002{\natexlab{a}}). \doi{10.1086/339442}

\bibitem[{\textbf{{Trujillo Bueno} et~al.}(2002{\natexlab{b}})\textbf{{Trujillo
  Bueno}, {Landi Degl'Innocenti}, {Collados}, {Merenda}, and {Manso
  Sainz}}}]{2002Natur.415..403T}
{Trujillo Bueno}, J., {Landi Degl'Innocenti}, E., {Collados}, M., {Merenda},
  L., and {Manso Sainz}, R., {Selective absorption
  processes as the origin of puzzling spectral line polarization from the Sun},
  \emph{\nat}, 415, 403 (2002{\natexlab{b}}).

\bibitem[{\textbf{{Trujillo Bueno} et~al.}(2004)\textbf{{Trujillo Bueno},
  {Shchukina}, and {Asensio Ramos}}}]{2004Natur.430..326T}
{Trujillo Bueno}, J., {Shchukina}, N., and {Asensio Ramos}, A., {A
  substantial amount of hidden magnetic energy in the quiet Sun}, \emph{\nat},
  430, 326 (2004). \doi{10.1038/nature02669}

\bibitem[{\textbf{{Vial} et~al.}(1980)\textbf{{Vial}, {Lemaire}, {Artzner}, and
  {Gouttebroze}}}]{1980SoPh...68..187V}
{Vial}, J.~C., {Lemaire}, P., {Artzner}, G., and {Gouttebroze}, P.,
  {\ion{O}{vi} ($\lambda=1032$~{\AA}) profiles in and above an active region
  prominence, compared to quiet sun center and limb profiles}, \emph{\solphys},
  68, 187 (1980). \doi{10.1007/BF00153276}

\bibitem[{\textbf{{White}}(1999)}]{1999SoPh..190..309W}
{White}, S.~M., {Radio Versus EUV/X-Ray Observations of the Solar
  Atmosphere}, \emph{\solphys}, 190, 309 (1999). \doi{10.1023/A:1005253501584}

\bibitem[{\textbf{{Wiegelmann} et~al.}(2014)\textbf{{Wiegelmann}, {Thalmann},
  and {Solanki}}}]{2014A&ARv..22...78W} {Wiegelmann}, T., {Thalmann}, J.~K., and {Solanki}, S.~K., {The magnetic field in the solar atmosphere}, \emph{\aapr}, 22, 78 (2014). \doi{10.1007/s00159-014-0078-7}

\bibitem[{\textbf{{Wilhelm} et~al.}(1995)\textbf{{Wilhelm}, {Curdt}, {Marsch} et~al.}}]{1995SoPh..162..189W}
{Wilhelm}, K., {Curdt}, W., {Marsch}, E., et~al., {SUMER - Solar Ultraviolet Measurements of
  Emitted Radiation}, \emph{\solphys}, 162, 189 (1995). \doi{10.1007/BF00733430}

\bibitem[{\textbf{{Withbroe} et~al.}(1982)\textbf{{Withbroe}, {Kohl}, {Weiser},
  and {Munro}}}]{1982SSRv...33...17W}
{Withbroe}, G.~L., {Kohl}, J.~L., {Weiser}, H., and {Munro}, R.~H.,
  {Probing the solar wind acceleration region using spectroscopic techniques},
  \emph{\ssr}, 33, 17 (1982). \doi{10.1007/BF00213247}

\bibitem[{\textbf{{Xu} et~al.}(2010)\textbf{{Xu}, {Lagg}, and
  {Solanki}}}]{2010A&A...520A..77X}
{Xu}, Z., {Lagg}, A., and {Solanki}, S.~K., {Magnetic structures of an
  emerging flux region in the solar photosphere and chromosphere}, \emph{\aap},
  520, A77 (2010). \doi{10.1051/0004-6361/200913227}


\bibitem[{\textbf{{Xu} et~al.}(2012)\textbf{{Xu}, {Lagg}, {Solanki}, and {Liu}}}]{2012ApJ...749..138X}
{Xu}, Z., {Lagg}, A., {Solanki}, S.~K., and {Liu}, Y., {Magnetic Fields of an Active Region Filament from Full Stokes Analysis of \ion{Si}{i} 1082.7 nm and \ion{He}{i} 1083.0 nm}, \emph{\apj}, 749, 138 (2012). \doi{10.1088/0004-637X/749/2/138}

\end{thebibliography}

\end{document}